\documentclass[reprint,prl]{revtex4-1}
\usepackage{bm}
\usepackage{graphicx}
\usepackage{amsmath}
\usepackage{hyperref}

\begin{document}
\title{Evolution of Band Topology by Competing Band Overlap
and Spin-Orbit Coupling: Twin Dirac Cones in Ba$_{\text{3}}$SnO as a Prototype}

\date{\today}
\author{Toshikaze Kariyado}
\affiliation{International Center for Materials Nanoarchitectonics,
National Institute for Materials Science, Tsukuba 305-0044, Japan}
\email{kariyado.toshikaze@nims.go.jp}
\author{Masao Ogata}
\affiliation{Department of Physics, the University of Tokyo, Bunkyo,
Tokyo 113-0033, Japan}
\begin{abstract}
 We theoretically demonstrate how competition between band inversion and
 spin-orbit coupling (SOC) results in nontrivial evolution of band
 topology, taking antiperovskite Ba$_{\text{3}}$SnO as a prototype
 material. A key observation is that when the band inversion dominates
 over SOC, there appear ``twin'' Dirac cones in the band structure. Due
 to the twin Dirac cones, the band shows highly peculiar structure in
 which the upper cone of one of the twin continuously transforms to the
 lower cone of the other. Interestingly, the relative
 size of the band inversion and SOC is controlled in this series of
 antiperovskite A$_3$EO by substitution of A (Ca, Sr, Ba) and/or E
 (Sn, Pb) atoms.
 Analysis of an effective model shows that the
 emergence of twin Dirac cones is general, which makes our
 argument a
 promising starting point for finding a singular band structure
 induced by the competing band inversion and SOC.
\end{abstract}

\maketitle

Singularities in band structures give rise to singular and attracting
responses of
materials
\cite{JPSJ.28.570,RevModPhys.81.109,Burkov:2016aa,Jia:2016aa,amv-review}. Therefore,
significant amount of efforts have been paid on
finding what kinds of singularities are possible in principle, and on
proposing materials to realize the singularities, both in
theory and experiments
\cite{0953-8984-28-30-303001,doi:10.1146/annurev-conmatphys-031016-025225,doi:10.1146/annurev-conmatphys-031016-025458}. The
``standard'' singularity is Dirac/Weyl cone
\cite{Wallace1947,Wan2011,BurkovWeyl2011,Wang2012,PhysRevB.88.125427},
which is characterized by a linear, or conical dispersion around an
isolated gap-closing point in the Brillouin
zone. Already a number of materials are confirmed to possess Dirac/Weyl
cones, ranging from graphene \cite{Novoselov:2005fk} (two-dimensional)
to Cd$_3$As$_2$ \cite{Neupane:2014aa},
Na$_3$Bi \cite{Liu2014}, and TaAs \cite{Lv:2015aa} (three-dimensional),
and so on.
However, the linear dispersion around the isolated gap-closing point
is not the only possible band singularity. The gap-closing point can
form a line (typically a loop) in the Brillouin zone rather than an
isolated point
\cite{PhysRev.52.365,Burkovnodal2011,doi:10.7566/JPSJ.85.013708}. Also,
a dispersion around a gap-closing point can be
quadratic rather than linear \cite{PhysRevLett.103.046811}. Very
recently, singularities involving three bands are also discussed
\cite{a161008877,PhysRevX.6.031003}.

Quite often, the three ingredients, i.e., band inversion, spin-orbit coupling
(SOC), and
symmetry, play key roles in generating a singular band structure. Roughly
speaking, the band inversion means an overlap between two bands with
different origin, typically originating from different kinds of
orbitals, say s- and p-orbitals, or p- and d-orbitals. Naively, the
overlapped bands repel with each other (band
anticrossing). However
the gap opening due to the band repulsion is sometimes prohibited by some
symmetry, leading to singular gap closing
points. The SOC, on the other hand, determines which kinds of the band
repulsion is allowed, in other words, what kinds of
gap-closing points can appear.

\begin{figure}[tb]
 \begin{center}
  \includegraphics{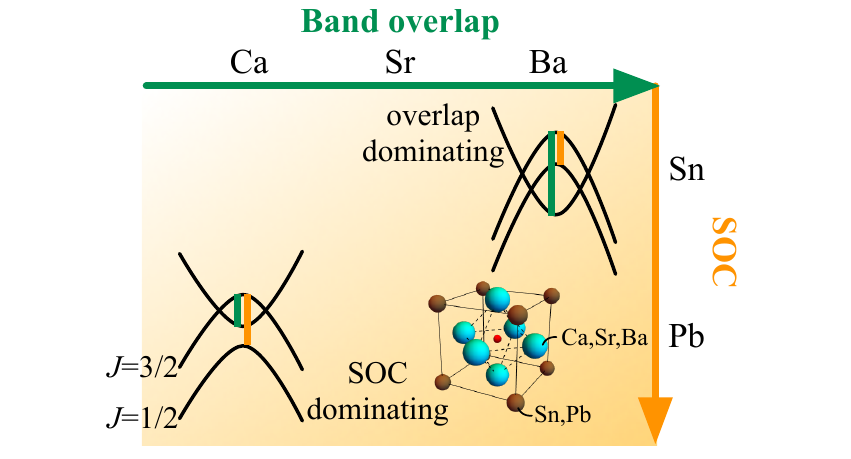}
  \caption{Schematic description of the evolution of the band topology
  as a function of the magnitude of the band overlap and the spin-orbit
  coupling. The crystal structure is shown as inset.}
  \label{fig:concept}
 \end{center}
\end{figure}
In this respect, antiperovskite A$_3$EO family (A=Ca, Sr, Ba and E=Pb,
Sn) is an ideal playground to observe singular band structures
\cite{JPSJ.80.083704,JPSJ.81.064701,klintenberg}. The representative
material Ca$_3$PbO is predicted to have
three-dimensional Dirac cones exactly at the Fermi energy,
and importantly, all of the three ingredients, the band
overlap, SOC, and the symmetry, are relevant in the formation of Dirac cones.
(Strictly speaking, there is a tiny mass gap. If we make an
emphasis on the gapful nature, the system is a topological crystalline
insulator \cite{PhysRevB.90.081112}.) It is also worth
noting that experiments on this series of materials are now
developing. We have highly refined crystal characterization
\cite{Nuss:dk5032}, thin film
fabrication \cite{doi:10.1063/1.4820770,doi:10.1063/1.4955213}, a thermal
measurement
\cite{doi:10.1063/1.4952393}, and even a report of superconductivity
\cite{Oudah:2016aa}.

In this paper, we show that the competition between SOC and the band
overlap leads to an interesting evolution of the band structure,
taking Ba$_3$SnO as a prototypical example. In the previous work on the
analysis of the antiperovskite family \cite{JPSJ.80.083704}, we regarded
that the SOC is dominant, and then, the Dirac cones appear at the Fermi
energy as mentioned above. In contrast, we will show that a new kind of
band singularity is generated when SOC is
smaller than the band inversion. The new singularity involves three (six
if we count the spin degrees of freedom) bands, which is similar to Refs.~\onlinecite{a161008877} and
\onlinecite{PhysRevX.6.031003}. Specifically, we find
``twin'' Dirac cones, which have small separation in the energy and
momentum space. The upper branch of the lower-energy Dirac cone
continuously transforms into the lower branch of the higher-energy Dirac
cone, which results in a peculiar evolution of the isoenergy surfaces as a
function of the energy. It is worth noting that the size of the
band overlap can be controlled by the choice of Ca, Sr, or Ba, while the size of the SOC
by the choice of Sn or Pb in A$_3$EO, namely, we can
design the band singularity. (See Fig.~\ref{fig:concept}.) We further
propose a simple
effective model to extract the
essence of this band singularity and show that the emergence of the twin
Dirac cones is general. These findings
form a concise guideline to search for novel states of matter originating
from the competition between SOC and the band overlap. 

Let us start with describing the band structure of
Ba$_3$SnO. The first-principles calculation is carried out with Quantum Espresso
package \cite{QE-2009}. In order to incorporate SOC, the relativistic
projector augmented wave (PAW) data sets from pslibrary 1.0.0 are
employed \cite{pslibrary}. The cutoff energies for the wave
function and the charge density are 50 and 400 Ry, respectively. 
We use the experimental crystal structure, i.e., cubic with the lattice
constant $a_0=5.448$ \AA~\cite{Widera19801805}. The recent refinement of
the crystal structure
reveals that the zero temperature structure is slightly distorted from the cubic
structure \cite{Nuss:dk5032}. However, since the symmetry of the crystal
plays an important
role, we stick with the cubic symmetry realized at room
temperature. Strictly speaking, the density functional theory for the
first-principles calculation is for zero temperature, but we
assume that the essence of the band singularity is safely captured. The
effect of the distortion will be briefly discussed at the end.

\begin{figure}[tb]
 \begin{center}
  \includegraphics[scale=1.0]{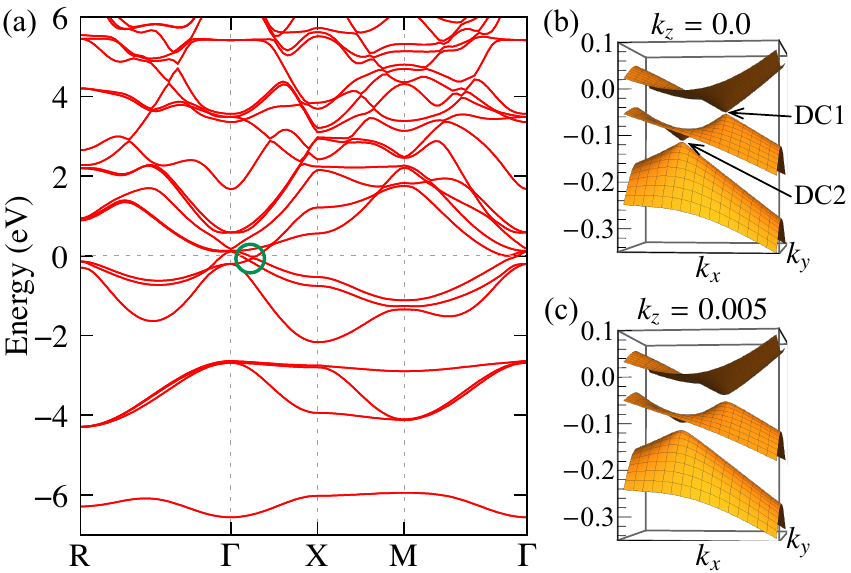}
  \caption{(Color online) (a) Global band structure of Ba$_{\text{3}}$SnO. (b,c)
  Dispersion relations of the three bands relevant to the twin Dirac
  cones. The plotted region is
  $0.07\leq k_x\leq 0.17$ and $-0.05\leq k_y\leq 0.05$ with $k_z=0.0$
  (b) and $kz=0.005$
  (c). The momenta are measured in the unit of $2\pi/a_0$, where $a_0$
  represents the lattice constant.}
  \label{fig:global}
 \end{center}
\end{figure}

The obtained band structure is shown in Fig.~\ref{fig:global}(a). The entangled
bands above the Fermi energy stem from d-orbitals (Ba-5d), while the
bands right below the Fermi energy from p-orbitals (Sn-5p). An
inspection of the orbital character of each
band reveals
that the top of the p-bands lies above the bottom of the d-bands, i.e.,
there is an overlap. There are two (twin) Dirac cones in
the region marked by a circle on $\Gamma$-X line in Fig.~\ref{fig:global}(a).
The singular structure of these twin cones becomes clear in the
three-dimensional plots of the dispersion relations shown in 
Figs.~\ref{fig:global}(b) and \ref{fig:global}(c). For $k_z=0$ [Fig.~\ref{fig:global}(b)],
we observe two Dirac cone structures
with small separation, and we denote the higher and lower energy ones DC1
and DC2, respectively. The upper branch of DC2 is
continuously turns into the lower branch of DC1.
Figure~\ref{fig:global}(c) with $k_z=0.005$ shows that the gap grows as $|k_z|$ increases,
which signals the three-dimensional nature of the Dirac cones. 
If we look at the band structure closely, it turns out that DC1 has a
tiny mass gap (less than 2 meV), while DC2 is exactly gapless. 
Note that the existence of the two Dirac cones on the 
$\Gamma$-X line implies that there are 12 Dirac cones in the entire
Brillouin zone due to the cubic symmetry.

\begin{figure}[tbp]
 \begin{center}
  \includegraphics[scale=1.0]{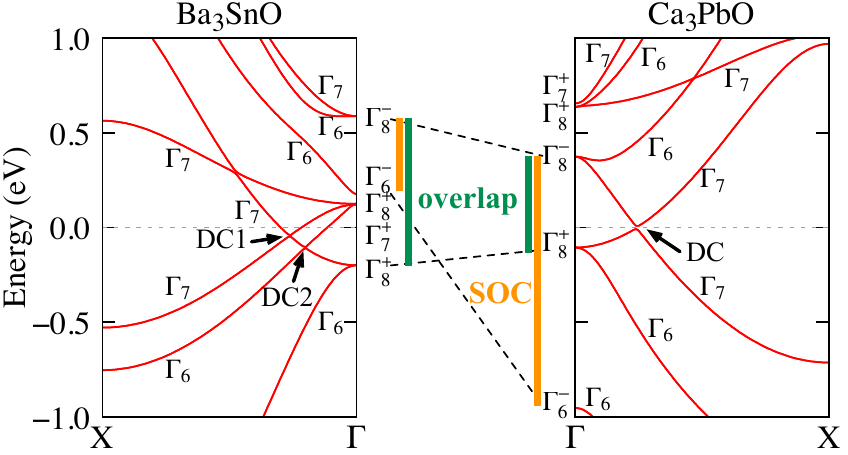}
  \caption{(Color online) Band structures on the $\Gamma$--X line along
  with the
  irreducible representations of the bands. The left (right) panel shows
  the band structure of Ba$_{\text{3}}$SnO (Ca$_{\text{3}}$PbO). The Dirac
  points are marked with arrows.}
  \label{fig:compare}
 \end{center}
\end{figure}

Let us compare Ba$_3$SnO and Ca$_3$PbO more in detail in order to
elucidate the
essence determining the difference in the band
topology. As in Fig.~\ref{fig:concept}, these
compounds share the same feature of the overlapping d- and p-bands, but
the critical difference lies in the magnitude of the
band overlap and SOC. We can see this from Fig.~\ref{fig:compare}, where the
close-up view of the band structure on $\Gamma$-X line is shown with
the irreducible representation for each band. Owing to SOC,
p-bands are split into $J=3/2$ and $J=1/2$ states. At
$\Gamma$-point, $\Gamma_8^-$ and $\Gamma_6^-$ correspond $J=3/2$ and
$J=1/2$ states, respectively. Therefore, the energy difference between
$\Gamma_8^-$ and $\Gamma_6^-$, represented as orange vertical
bars in Fig.~\ref{fig:compare}, measures the size of SOC. On the other
hand, in the energy range of Fig.~\ref{fig:compare}, the lowest-energy
$\Gamma_8^+$
state at $\Gamma$-point represents the bottom of the d-bands, and thus, the
energy difference between the lowest $\Gamma_8^+$ and $\Gamma_8^-$
(green bars in Fig.~\ref{fig:compare}) is regarded as the size of the
band overlap. Apparently, the band overlap is larger than SOC in
Ba$_3$SnO, while it is opposite in Ca$_3$PbO. 

Using this difference in the size of SOC and band overlap, the band
topologies in Ba$_3$SnO and Ca$_3$PbO are 
explained as follows. By definition, when the band overlap is larger
than SOC as in Ba$_3$SnO, the $J=1/2$ band is above the bottom of the d-bands at
$\Gamma$-point, while when SOC is larger than the band overlap as in
Ca$_3$PbO, the $J=1/2$ band comes below the bottom of the d-bands. Then, by following
the way in which the bands are connected in Fig.~\ref{fig:compare}, we see that the gap-closing point defined
as DC2 is allowed only when the $J=1/2$ band is above the bottom of the
d-bands as in the case of Ba$_3$SnO. Therefore, two Dirac cones, DC1 and
DC2 appear in Ba$_3$SnO, while only one in Ca$_3$PbO. Next, the reason
for
Ba$_3$SnO having larger band overlap than Ca$_3$PbO is that the band
width of the Ba-5d orbital is larger than that of the Ca-3d orbital, and
as a result, the bottom of the Ba-5d band is lower than that of the
Ca-3d band. This difference between Ba-5d and Ca-3d originates from 
the difference of the spreading of the wave functions for Ba-5d 
and Ca-3d orbitals. On the other hand, SOC in this series of materials
is mostly provided by Sn or Pb atoms. Therefore SOC is weaker in
Ba$_3$SnO than Ca$_3$PbO. To summarize, Ba$_3$SnO is a
band-overlap-dominant material whereas Ca$_3$PbO is an SOC-dominant
material, which explains the difference in the band topology. The
argument here suggests that the band topology can be controlled by
partial substitution of A and E atoms in A$_3$EO. (See Fig.~\ref{fig:concept}.)

From Fig.~\ref{fig:compare}, we can also understand the existence and
nonexistence of the mass gap at DC1 and DC2. We see that DC1
is a crossing between two $\Gamma_7$ bands, while DC2 is a crossing
between $\Gamma_6$ and $\Gamma_7$. Therefore, DC1 can have finite mass
gap in principle. Specifically, the gap at DC1 is induced by the
indirect spin-orbit coupling involving the states far away from the
Fermi energy as discussed before in Ca$_3$PbO
\cite{JPSJ.80.083704,JPSJ.81.064701}. Since it involves the
far away states, the gap at DC1 is tiny as obtained in the
first-principles calculation. In contrast, DC2 is exactly gapless, which is
protected by the crystalline symmetry, just as in the case of the known
Dirac semimetals like Cd$_3$As$_2$ or Na$_3$Bi. 

\begin{figure}[tbp]
 \begin{center}
  \includegraphics[scale=1.0]{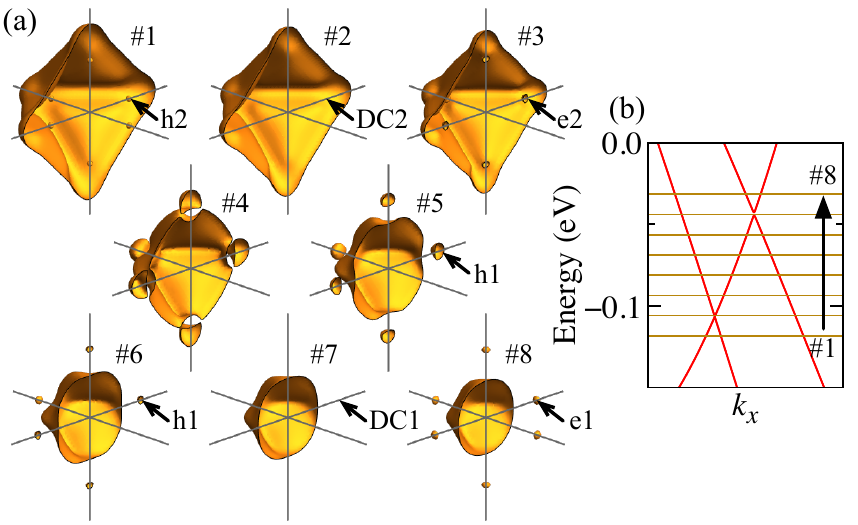}
  \caption{(Color online) Energy evolution of the isoenergy surfaces of
  the three bands relevant to the Dirac cones. The selected energy for
  each surface is represented in (b). In each panel, some part of the
  isoenergy surface is eliminated to make the inside visible.}
  \label{fig:iso}
 \end{center}
\end{figure}

A highly peculiar band structure associated with the twin
Dirac cones becomes apparent if we look at the isoenergy surfaces for several
energies, which would represent the doping dependence of the Fermi
surfaces in the rigid-band assumption. Figure~\ref{fig:iso} shows the
isoenergy surfaces for energies \#1
$-0.1184$, \#2 $-0.106$, \#3 $-0.0936$, \#4 $-0.0812$, \#5 $-0.0688$,
\#6 $-0.0564$, \#7 $-0.044$, and \#8 $-0.0316$ in the unit of eV measured from
the undoped Fermi energy. [Figure~\ref{fig:iso}(b) shows the selected energies.] In \#1, we observe two types of
isoenergy surfaces: one is a large surface associated with the
non-Dirac hole-like band, and the others are small and almost point-like
surfaces [shown as h2 in Fig.~\ref{fig:iso}(a)] associated with the lower branch of
DC2 inside of the larger surface. 
As the energy is increased, i.e. \#1$\rightarrow$\#3,
the large hole-like surface gradually shrinks. On the other hand, h2
first shrinks, disappears at \#2 (Dirac point of DC2), and reappears as
a small
surface e2,
which is the upper branch of DC2 having electron-like nature. As the
energy increases further, e2 expands, and at a certain energy around
\#4, topology of the isoenergy
surfaces changes. After this topological change, e2 turns into h1,
which is now hole-like and outside of the larger surface. Then,
there appear small hole surfaces (h1) as shown in \#5.
Since h1 is the lower branch of DC1, it 
disappears at \#7 (Dirac point of DC1) and reappears as a small
electron-like surface e1 when
the energy is
further increased. The discovered topological change of the isoenergy
surface must cause interesting doping dependent responses in
measurements sensitive to Fermi surface shapes, such as the quantum
oscillation.

Next, we show that the emergence of the twin Dirac cones are generic by
constructing a minimal effective model. For this purpose, we focus on
the band structure along the $k_z$-axis, on which
$C_{4v}$ point-group symmetry is effective. 
Before taking SOC into account, there are five irreducible
representations in $C_{4v}$: A$_1$, A$_2$, B$_1$, B$_2$, and E.
Intuitively, $s$, $p_z$, and
$d_{3z^2-r^2}$ orbitals correspond to A$_1$ representation,
$d_{x^2-y^2}$ orbital to B$_1$,
$d_{xy}$ orbital to B$_2$, and
$p_{x/y}$ and $d_{zx/yz}$ orbitals to E. 
From the orbital weight analysis of the bands, we find 
that $p_{x/y}$ orbitals of Sn (or Pb) and $d_{x^2-y^2}$ orbitals of Ba
(or Ca, Sr) are responsible for the twin generation. Therefore, we keep
the
E and B$_1$ states in the minimal effective model. By adding SOC as
$\bm{L}\cdot\bm{S}$ coupling on the atomic p-orbitals, we obtain the
effective Hamiltonian as
\begin{equation}
 \hat{H}=
  \setlength{\arraycolsep}{0.5mm}
  \begin{pmatrix}
   g_p(k_z)&-\mathrm{i}\lambda &hk_x&0&0&0\\
   \mathrm{i}\lambda&g_p(k_z) &-hk_y &0 &0 &0 \\
   hk_x&-hk_y &g_d(k_z) &0 &0 &0\\ 
   0&0 &0 &g_p(k_z) &\mathrm{i}\lambda & hk_x\\ 
   0&0 &0 &-\mathrm{i}\lambda &g_p(k_z) &-hk_y\\
   0&0 &0 &hk_x &-hk_y & g_d(k_z)
  \end{pmatrix},\label{H0}
\end{equation}
where the basis wave functions are $|p_x\uparrow\rangle$,
$|p_y\uparrow\rangle$, $|d\uparrow\rangle$, $|p_x\downarrow\rangle$,
$|p_y\downarrow\rangle$, $|d\downarrow\rangle$, and 
$\lambda$ is the coefficient for SOC.
The matrix elements in Eq.~\eqref{H0} are kept up to the first order
in $k_x$ and $k_y$, since we focus on the effective model in the
vicinity of $k_z$-axis. Due to the difference in symmetry between $p_{x/y}$
and $d_{x^2-y^2}$ orbitals, the offdiagonal elements are linear in $k_x$
and $k_y$.

When we switch to the basis set that diagonalizes SOC, the 
Hamiltonian Eq.~\eqref{H0} becomes
$\hat{H}=\hat{H}_\lambda\oplus\hat{H}_{-\lambda}$ with
\begin{equation}
 \hat{H}_\lambda=
  \begin{pmatrix}
   g_p(k_z)+\lambda&0&hk_-\\
   0&g_p(k_z)-\lambda &hk_+\\ 
   hk_+&hk_- & g_d(k_z)
  \end{pmatrix},\label{H1} 
\end{equation}
where $k_\pm\equiv k_x\pm \mathrm{i}k_y$. 
Now, suppose that there exists $k_{z0}^{\pm}$ satisfying
$g_p(k_{z0}^\pm)\pm\lambda=g_d(k_{z0}^\pm)$. Then, near
$k_z=k_{z0}^\pm$, $g_p(k_z)$ and $g_d(k_z)$ can be expanded as
$g_p(k_z)\pm\lambda=\epsilon_\pm+c_{p}^\pm(k_z-k_{z0}^\pm)$ and
$g_d(k_z)=\epsilon_\pm-c_d^\pm(k_z-k_{z0}^\pm)$. Using this expression,
the effective Hamiltonian around $\bm{k}=(0,0,k_{z0}^\pm)$ becomes
$\hat{H}=\hat{H}^{(+)}\oplus\hat{H}^{(-)}$ with
\begin{equation}
 \hat{H}^{(\pm)}=
  (\epsilon_{\pm}+\delta c_\pm\delta k_z^{\pm})+
  \begin{pmatrix}
   c_\pm\delta k_z^\pm& hk_\mp \\
   hk_\pm& -c_\pm\delta k_z^\pm
  \end{pmatrix},\label{Hfinal}
\end{equation}
where $\delta k_z^{\pm}=k_z-k_{z0}^\pm$, $c_\pm=(c_p^\pm+c_d^\pm)/2$, and
$\delta c_\pm=(c_p^\pm-c_d^\pm)/2$. This effective Hamiltonian is
nothing more than Dirac
Hamiltonian. The parameter $\delta c_\pm$ in Eq.~\eqref{Hfinal}
determines the tilting angle of the Dirac cone, and when the slopes of
$g_p(k_z)$ and $g_d(k_z)$ are nearly opposite (i.e., $c_p^\pm\sim
c_d^\pm$), this tilting angle becomes small.  Note
that the state neglected in deriving Eq.~\eqref{Hfinal} from
Eq.~\eqref{H1} only affects the resultant model at least in the second
order with respect to $k_\pm$. 

The above results are interpreted in the language of the competing band
overlap and SOC as follows. Firstly, for the band-overlap dominant case,
e.g., in Ba$_3$SnO,
both of the two conditions, $g_p(k_z)+\lambda=g_d(k_z)$ and
$g_p(k_z)-\lambda=g_d(k_z)$ can be satisfied, which leads to the twin
Dirac cones. On the other hand, for the SOC dominant case, e.g., in
Ca$_3$PbO, only $g_p(k_z)+\lambda=g_d(k_z)$ has a
valid solution, while $g_p(k_z)-\lambda=g_d(k_z)$ cannot have a solution
at any $k_z$ in the Brillouin zone since $\lambda$ is large. In this
way, the number
of Dirac cones is correctly reproduced in this minimal effective model. It
is also worth noting that the twin Dirac cones are merged in the $\lambda=0$
limit
since two conditions $g_p(k_z)\pm\lambda= g_d(k_z)$ become
identical. Inversely speaking, the twin cones are generated by
SOC. Although the above model is derived with a specific kind of
material in
mind, the strong restriction is only from the cubic symmetry, and thus,
the model can be realized in a wide class of materials. All we have to
do is to find a way to control the band-overlap and SOC, which is
relatively simple in our A$_3$EO by partial
substitution in A and E atoms. 

Before closing, we make a few comments on our theoretical
approach. Firstly, we have considered the room temperature cubic
structure
instead of the zero temperature distorted structure, in order to take
advantage of the high symmetry. We have performed some
band structure calculation with distortion (not shown). We find that, 
because of the symmetry reduction, even DC2 becomes gapful, although the
gap is small partially because the distortion is still
perturbative to the original cubic structure \cite{Nuss:dk5032}. The
second comment is about our
computational methods. We have applied a standard density functional
theory, but we should note that a precise prediction of the band
overlap size is challenging \cite{PhysRevB.84.041109}. For example, HSE correction
gives different numbers for the size of the overlap
\cite{PhysRevB.90.081112}. However,
the overall tendency is unaltered by some
computational details, i.e., the tendency that Ba based
compound has a larger band overlap than the Ca based one will not
change. Therefore, the essence of the present paper to give a unified
view on the band topology due to the competition between SOC and band
overlap will be unchanged.

To summarize, the band structure of Ba$_3$SnO is investigated by means of
the first-principles method, and twin Dirac cones and associated
peculiar band dispersions are disclosed. Behind this phenomenon, there is a
competition between the band overlap and the spin-orbit
coupling. Further study on A$_3$EO (A=Ca,Sr,Ba and E=Sn,Pb) should be
worthwhile, since it provides a simple way to control the band overlap
and the spin-orbit coupling by chemical substitution. We believe that
the concept of competing band overlap and spin-orbit coupling is useful
in future exploration of novel states of matter. 

\begin{acknowledgments}
The authors thank H.~Takagi for stimulating discussions. TK thanks the
Supercomputer Center, the Institute for Solid State Physics, the
 University of Tokyo for the use of the facilities.
 This work was supported by a Grant-in-Aid for Scientific Research
 No.~15H02108 (MO) and No.~17K14358 (TK).
\end{acknowledgments}

\end{document}